\documentstyle[prl,aps,preprint]{revtex}
\begin{document}
\tighten

\title{ONE-LOOP QCD CORRECTIONS TO \\ DEEPLY-VIRTUAL COMPTON
SCATTERING: \\ THE PARTON HELICITY-INDEPENDENT CASE
}
\author{Xiangdong Ji and Jonathan Osborne}
\bigskip

\address{
Department of Physics \\
University of Maryland \\
College Park, Maryland 20742 \\
{~}}

\date{UMD PP\#97-001 ~~~DOE/ER/40762-124~~~ July 1997}

\maketitle

\begin{abstract}
We show that the one-loop QCD correction to deeply-virtual
Compton scattering can be factorized into the finite 
perturbative contributions and the collinearly-divergent
terms, which correspond to the matrix elements 
of the off-forward parton distributions. As a by-product, we
obtain the next-to-leading order coefficient functions
in the generalized operator product expansion of two 
vector currents. 

\end{abstract}
\pacs{xxxxxx}

\narrowtext

In searching for ways to measure the amount of the nucleon
spin carried by quark orbital angular momentum, one 
of us introduced deeply-virtual compton scattering
(DVCS) as a probe to a novel class of 
``off-forward" parton distributions (OFPDs)
\cite{ji1}. DVCS is a process in which a highly virtual
photon (with virtual mass $Q^2>\!\!>\Lambda_{\rm QCD}^2$)
scatters on a nucleon target (polarized or unpolarized),
producing an exclusive final state consisting of a
high-energy real photon plus a slightly recoiled 
nucleon. With the virtual photon in the Bjorken limit, 
a quantum chromodynamic (QCD) analysis shows that the scattering
is dominated by the simple mechanism in which a quark
(antiquark) in the initial nucleon absorbs the virtual photon, 
immediately radiates a real one, and falls back to 
form the recoiled nucleon. 

Several interesting theoretical papers have 
since appeared in the literature, which made further
studies of the DVCS process \cite{ra1,ji2,ra2,chen,di} 
and the OFPDs 
\cite{ji3,bl,bal}. A process closely-related to DVCS,
Compton scattering with two-virtual photons, have
been studied before from theoretical interests
\cite{wana,muller1}. The OFPDs
have also been recognized in a number of
theoretical studies in the past\cite{geyer,ralston}.
Moreover, the QCD evolution equations for the OFPDs
\cite{ra1,ji2,bl,bal,muller1}
are closely related to the evolution of meson wave
functions and light-ray or string operators \cite{all}. 
However, physical significance and actual 
measurement of OFPDs have not been
thoroughly explored in the literature. It was pointed out 
in Ref. \cite{ji1} that the OFPDs can appear in 
general types of hard diffractive processes. Recent works 
in Ref. \cite{ra3,hoodbhoy,collins,frankfurt} have indeed found 
their use in diffractive meson production 
\cite{brodsky,frankfurt1} and $Z^0$ or muon pair production 
\cite{bartels,gribov}.

In this paper, we study the one-loop 
correction to DVCS in QCD.
Here, for simplicity, we consider only the parton
helicity-independent amplitude. [The helicity-dependent
case will be published separately, together with
a more detailed analysis of the present calculation \cite{os}.]
Our result shows
that the infrared divergences in the one-loop
diagrams can be entirely factorized into the 
nonperturbative OFPDs. This result is not immediately obvious from
an analogy with a general virtual Compton scattering,
because in DVCS the final state 
photon is real and new infrared 
divergences can potentially arise from the 
appearance of a new light-like 
momentum. It turns out, however, that
a factorization theorem does hold at one-loop 
level with the final state photon real and 
structureless. We expect the factorization theorem
to be true to all orders in perturbation theory, 
allowing DVCS to be studied in perturbative QCD just like other 
gold-plated examples such as deep-inelastic scattering.
The finite part of our one-loop results, together
with the two-loop evolution of OFPDs in the $\rm \overline{MS}$
scheme, provides the necessary ingredients for calculating 
DVCS at the next-to-leading order.

To carry out our study in a more general setting,
we actually consider the non-forward,
unequal-mass virtual Compton scattering 
\cite{chen,wana,muller1}.
We call the incoming (out-going) virtual photon momentum 
$q^\mu$ ($q'^\mu=q^\mu-\Delta^\mu$), and the incoming (out-going)
nucleon momentum $P^\mu$ ($P'^\mu=P^\mu+\Delta^\mu$). We introduce 
the ``average" photon momentum $\bar q^\mu=(q+q')^\mu/2$ and
``average" proton momentum $\bar P^\mu = (P+P')^\mu/2$. 
The Compton amplitude is defined as,
\begin{equation}
   T^{\mu\nu} = i\int d^4z
     e^{-i\bar q\cdot z}\left\langle P'\left|TJ^\mu\left(-{z\over 2}
           \right) J^\nu \left({z\over 2}\right)
       \right|P\right\rangle \ . 
\end{equation}
We want to study the simplification of
the amplitude in the Bjorken limit: ${\bar Q}^2=-{\bar q}^2\rightarrow
\infty$, $\bar P\cdot \bar Q\rightarrow \infty$,
and the ratio of the two staying finite.

For convenience, we choose two light-like momenta $p^\mu$ and $n^\mu$
so that $p^2=n^2=0$ and $p\cdot n=1$. We let the average
photon and nucleon momenta to be parallel to $p^\mu$
and $n^\mu$, so that, 
\begin{eqnarray}
    \bar P^\mu &=& p^\mu + {\bar M^2\over 2} n^\mu \ , \nonumber \\
     \bar q^\mu &=& -x_B p^\mu + {\bar Q^2\over 2x_B} n^\mu \ ,  
\end{eqnarray}
where $\bar M^2 = M^2 - \Delta^2/4$ and $M$ is the mass of the nucleon.
The Bjorken $x_B$ is ${\bar Q}^2/(2\bar P\cdot \bar q)$,
where corrections of order $1/{\bar Q}^2$ are ignored.
Clearly, any four-momentum can be expanded in terms
of $p^\mu$, $n^\mu$, and two unit vectors in the
transverse directions. Large scalars in any Feynman diagram 
come from the product of the $p$ component of a four-vector 
and the $n$ component of $\bar q$. Therefore, in the
leading order, we can safely ignore 
other components of an external momentum. 
For instance, the momenta of the initial and final state 
nucleons can be approximated 
as $(1+\xi)p^\mu$ and $(1-\xi)p^\mu$, respectively, 
where $\xi$ is defined
form the expansion,
\begin{equation}
    \Delta^\mu = -2\xi \ p^\mu + ... \ . 
\end{equation}
and is constrained to [0,1] by our choice of coordinates. [Note
that the definition of $\xi$ differs from that in Ref. \cite{ji1}
by a factor of 2.] 
Moreover, the momenta of initial and final partons 
participating in a hard subprocess can effectively
be taken as $(x+\xi)p^\mu$ and $(x-\xi)p^\mu$, respectively, 
where $-1<x<1$. 
The longitudinal momentum fractions of the initial and final nucleons 
carried by partons are $(x+\xi)/(1+\xi)$ and $(x-\xi)/(1-\xi)$, 
respectively.  

We consider only the symmetric part
of $T^{\mu\nu}$, which is insensitive to 
helicities of partons in a nucleon target. Like the
forward Compton scattering, there
are two leading-order Lorentz-invariant amplitudes \cite{chen},
\begin{equation}
     T^{\mu\nu} = (-g^{\mu\nu} +p^\mu n^\nu + p^\nu n^\mu)T_1
      + \left(p^\mu-{q'^\mu p\cdot q'\over q'^2}\right)
    \left(p^\nu-{q^\nu p\cdot q\over q^2} \right) T_L \ , 
\end{equation}
where $T_i$ are functions of 
$x_B$, $\xi$, $t=\Delta^2$, and 
${\bar Q}^2$. Physically, $T_1$ represents
the amplitude for the transversely-polarized 
photon scattering, and $T_L$ the amplitude for
the longitudinally-polarized scattering. 
Since $T_L$ does not
contribute to DVCS, we 
ignore it in the remainder of the paper. 

A factorization theorem is believed to exist for 
the general virtual Compton amplitudes defined above. For instance,
\begin{equation}
  T_1(x_B,\xi,t,\bar Q^2) = 
   \int^1_{-1} {dx\over x} \sum_a F_a(x,\xi,t,\bar Q^2)
    C_{1a}\left({x\over x_B},{\xi\over x_B}, \alpha_s(\bar Q^2)\right) \ , 
\label{fac}
\end{equation}
where $a$ labels different parton species. 
$F_a$ is the helicity independent, 
leading-twist off-forward parton distributions
as defined in Ref. \cite{ji1}. 
For quarks, one has,
\begin{equation}
      F_{a=q}(x) = {1\over 2}\int 
       {d\lambda\over 2\pi} e^{i\lambda x} \left\langle P'\left|\bar \psi
      \left(-{\lambda\over 2}n\right) \not\! n
         \psi\left({\lambda\over 2}n\right)\right| P\right\rangle \ , 
\end{equation}
and similarly for gluons, $F_G(x)$.  
$C_{1a}$ is the coefficient function calculable in perturbative 
QCD. At the leading order 
in $\alpha_s$, one has
\begin{equation}
     C_{1a}^0(x,\xi) = - e_a^2\left({x\over x-1} + {x\over x+1}\right) \ , 
\end{equation}
where $e_a$ is the electric charge.
The second term in the bracket represents a 
crossing contribution which is always present in
Compton processes. To simplify the presentation,
we will not include it explicitly in the following
formulas. For $\xi= t=0$, we go back to the 
well-known forward virtual Compton scattering.

The factorization formula works for both $x_B>1$ and $x_B<1$ 
regions. In fact, it is defined throughout the complex
plane of $x_B$. The amplitude is analytic at $|x_B|>1$,
where there is no on-shell propagation.
Therefore, one can expand $C_{1a}$ in Taylor series 
at $x_B=\infty$, 
\begin{equation}
   C_{1a}({x\over x_B},{\xi\over x_B}) 
= \sum_{n=2,4...~i=0,2...}^\infty C_a^{ni}(\alpha_s(\bar 
 Q^2)) {x^{n}  \xi^i \over x_B^{n+i}}
\end{equation}
The $x$-integration in Eq. (\ref{fac}) can now 
be done using 
\begin{equation}
\int^1_{-1} dx x^{n-1}F_a(x,\xi,t) = a_n(\xi,t) ={1\over 2} 
 n^{\mu_1}n^{\mu_2} \cdots n^{\mu_n}
        \langle P'|\overline \psi i\stackrel{\leftrightarrow}{D}^{\mu_1} \cdots i\stackrel{\leftrightarrow}{D}^{\mu_{n-1}}
        \gamma^{\mu_n} \psi |P\rangle  .
\end{equation}
Therefore the $i=0$ terms in the amplitude $T_1$ are 
just the result of the usual twist-2 
operators in the operator product expansion (OPE)
for forward virtual Compton scattering.
As for the $i\neq 0$ terms, one can translate the
$\xi$ factors into total derivatives on the
twist-2 operators,
\begin{equation}
   n_{\mu_1}...n_{\mu_{n+i}}  \langle P'|i\partial^{\mu_{n+1}}
       ...i\partial^{\mu_{n+i}}
      \overline \psi i\stackrel{\leftrightarrow}{D}^{\mu_1} \cdots i
   \stackrel{\leftrightarrow}{D}^{\mu_{n-1}}
        \gamma^{\mu_n} \psi |P \rangle  
    = 2(2\xi)^i a_n(\xi,t) \ . 
\end{equation}
Therefore the factorization formula in Eq. (\ref{fac})
reflects a general OPE in terms
of twist-2 operators with arbitrary numbers 
of total derivatives. $C^{n0}(\alpha_s(\bar Q^2))$ 
is the coefficient function relevant for
deep-inelastic scattering \cite{bardeen}. 
A general virtual Compton process also requires 
the $i\neq 0$ terms.

Deeply virtual Compton scattering is a physical process with the kinematic 
requirement that the final photon be on its mass shell. 
This corresponds to the region $x_B = \xi < 1$, 
where the OPE form
of the amplitude is not particularly illuminating. 
In particular, the Compton amplitudes
now have both real and imaginary parts. 
The analytic continuation to the $|x_B|<1$ region
is made by approaching the real axis from the lower
half plane, and hence 
$x_B$ has a small negative imaginary part.
The central question we would like to address 
is this: Does the factorization theorem
still hold when $x_B=\xi$?  One can, of course, study
this question using the  
general factorization techniques\cite{sterman}. While we believe the answer is 
affirmative, we present 
a one-loop calculation to support this.

We have calculated the general one-loop 
virtual Compton scattering with on-shell
quarks and gluons. We present the quark 
result first. The initial and final 
momenta of the quark target are taken to be 
$P^\mu=(x+\xi)p^\mu$ and $P'^\mu=(x-\xi)p^\mu$,
respectively. For convenience, we 
replace the spinor-space matrix element 
$\bar u(P')\Gamma u(P)$ by a trace $\rm {Tr}[\not\! p\Gamma]/2$.
Although they are not identical, this replacement
only affects the interpretation of the result. 
We call the resulting Compton amplitude, $t^{\mu\nu}$, 
with Lorentz-scalar 
amplitudes $t_i$. At the leading order, one
has,
\begin{equation}
      t_{1a} = -{e_a^2\over x-x_B + i\epsilon} \ .  
\end{equation}
A practical aspect of parton
calculation lies in the fact that the coefficient 
function $C_{1a}(x,\xi)$ can be obtained from the 
infrared subtracted parton scattering amplitude.  

We use dimensional regularization to regularize the 
infrared divergences present in the loop calculations
($d=4+\epsilon, \epsilon>0$). 
We use the Lehmann-Symanzik-Zimmermann reduction formula 
to extract on-shell matrix elements from the 
corresponding Green's functions. The calculation 
is done in the region where $x_B>1$ so that there is 
no imaginary part. Summing up the standard
vertex, self-energy, and box diagrams,
we find the following divergent contribution,
\begin{eqnarray}
t_{1a}^{\rm pole} &=& e_a^2 {\alpha_s\over 2\pi}C_F \left(-{2\over \epsilon}\right)\ 
 {1\over x-x_B}\left[ {3\over 2}
  + \frac{x^2 + x_B^2-2\xi^2}{x^2-\xi^2}\int^x_\xi {dy\over y-x_B}\right.
      \nonumber \\
  && \left.~~~~~~~ + \frac{(x_B+\xi)(x-x_B+2\xi)}{2\xi(x+\xi)}
      \int^\xi_{-\xi} \frac{dy}{y-x_B}\right] \ .  
\end{eqnarray}
An examination shows that the above 
expression is proportional  
to the perturbative matrix elements of the
OFPDs calculated
in dimensional regularization at one-loop order \cite{ji2}. 
Therefore, the effect of this term is already taken 
into account in the general factorization formula 
with a leading-order coefficient function 
$C_{1a}^0$ and the nonperturbative OFPDs. In the limit
$x_B\rightarrow \xi$, the factorization 
is not affected.

We also find the finite part of the one-loop result,
\begin{eqnarray}
 &&  t_{1a}^{\rm finite} = e_a^2{\alpha_s\over 2\pi} C_F 
       \left\{ -{9\omega \over 2(1-\omega x)} \right. \nonumber \\
   &&+\left[{3x(1-\omega^2\xi^2)\over (x^2-\xi^2)(1-\omega^2x^2)}
      - {x(1-\omega \xi)\over (x^2-\xi^2)}
        \left({1\over 2\omega\xi} + {1+\omega\xi\over 1-\omega^2x^2}
   \right) \log(1-\xi \omega)\right]\log(1-\omega\xi) \nonumber \\
   &&+ \left. \left[{(1-\omega x)^2 + 2(\omega x-\omega^2\xi^2)
     \over 2\omega(x^2-\xi^2)(1-\omega x)}\log(1-\omega x)
     -{3\over 2}{2x-\omega(x^2+\xi^2)\over (x^2-\xi^2)
       (1-\omega x)}\right]\log(1-\omega x) \right\} \ , 
\end{eqnarray}
where we have introduced $\omega = 1/x_B$. 
The above expression is manifestly finite in the DVCS limit,
$\omega \xi\rightarrow 1$, so the one-loop 
factorization theorem for DVCS holds for
quark scattering. When $\xi=0$, we recover the coefficient function for
the forward Compton scattering pertinent to 
deep-inelastic scattering \cite{bardeen}.
The coefficient function coupling with the off-forward 
quark distributions
in the general virtual Compton scattering at order 
$\alpha_s$ is,
\begin{equation}
       C_{1a}^1(x,\xi) = x t_{1a}^{\rm finite}(x, \xi,\omega=1) \ . 
\end{equation}

Next, we consider virtual Compton scattering on an on-shell gluon.
Again for convenience, we replace the polarization 
product $\epsilon^{\mu *}\epsilon^\nu$ by $(-g^{\mu\nu} + p^\mu
n^\nu + p^\nu n^\mu)/(2+\epsilon)$. We neglect the last two terms because
of the color gauge invariance. At one-loop level, there
are six quark-box diagrams, three of which have identical results
as the other three. Two of the three inequivalent
diagrams are related to each other
by crossing symmetry; and the remaining one is 
itself crossing-symmetric. 
The infrared divergent part of the amplitude is,
\begin{eqnarray}
  t_{1g}^{\rm pole} &=& \ {\alpha_s\over 2\pi} \ {2T_F\sum_a e_a^2}\ \left(
       - \frac{2}{\epsilon} \right)  
       \left[\frac{2xx_B}{x^2-\xi^2}
       + \frac{x_B^2+(x-x_B)^2-\xi^2}{x^2-\xi^2}
      \int^x_\xi {dy\over y-x_B} \right. \\ 
    &&\left.~ ~~~~~~ + \left(\frac{(x_B+\xi)(x-2x_B+\xi)}
     {2\xi(x+\xi)} + \frac{x(x_B^2-\xi^2)}{\xi(x^2-\xi^2)}\right)
       \int^\xi_{-\xi} \frac{dy}{y-x_B}\right]\ , 
\end{eqnarray}
where $T_F=1/2$. 
It is not immediately clear that all the terms above can be absorbed 
into the renormalization mixing of the quark distributions with
the gluon. However, notice that the helicity-independent 
gluon distribution $F_G(x)$ is antisymmetric in $x\rightarrow
-x$. Therefore, when convoluted with $F_G(x)$
only the $x$-symmetric part of the amplitude contributes. 
And the $x$-symmetric $t_{1g}^{\rm pole}$ is identical to 
the mixing coefficient in a renormalized
quark density $\cite{ji2}$.

The remaining finite part is considered as the gluon 
contribution to the Compton scattering
in the $\rm \overline{MS}$ scheme. Our calculation
yields,
\begin{eqnarray}
   t_{1g}^{\rm finite} &= &\frac{\alpha_s}{2\pi} (2T_F\sum_a e_a^2) \ 
            \left\{
\left[\left(1+ {2(1-\omega x)\over \omega^2(x^2-\xi^2)}
  \right)\left(1-{1\over 2}\log(1-\omega x)\right) \right. \right. 
   \nonumber \\
 &&  +\left. 2{ 1-\omega x\over \omega^2(x^2-\xi^2)}\right]\log(1-\omega x)   
    + \left[\left({1\over \omega \xi}
  -{2\over \omega^2(x^2-\xi^2)}\right) \right. \nonumber \\ 
 && \times \left. \left. \left(1-{1\over 2}
  \log(1-\xi\omega)\right)-{2\over \omega^2
(x^2-\xi^2)} \right](1-\omega\xi) \log(1-\xi\omega) \right\} \ . 
\end{eqnarray}    
The above expression is finite in the limit of 
$\omega\xi\rightarrow 1$, so the factorization for
DVCS on a gluon target holds at one-loop order. 
For $\xi=0$, our result agrees 
with the coefficient function 
calculated in \cite{bardeen}.
Using the definition of the off-forward gluon 
distribution $F_G(x)$, we found the coefficient 
function in the general factorization formula
at the $\alpha_s$ order, 
\begin{equation}
      C_{1g}^1(x, \xi) = {x^2\over x^2-\xi^2}t_{1g}^{\rm finite}(x, 
      \xi,\omega=1) \ . 
\end{equation}

In a recent paper, M$\rm \ddot{u}$ller \cite{muller}
studied the constraints of the spacetime conformal symmetry 
on the form of the OPE for two vector currents. He found that
at the next-to-leading order, the OPE can be expressed
in terms of the coefficient functions in the forward
scattering ($\xi=0$) and the improved conformal-covariant 
operators at the leading order. While the result can be 
checked for the non-singlet case using the two-loop
anomalous dimensions in the literature
\cite{twoloop}, such anomalous dimensions 
for the singlet case are not yet available. 
Clearly, they are needed if
one is interested in carrying the analysis for DVCS
to the full next-to-leading order.

\acknowledgements
We thank I. Balitsky, P. Hoodbhoy and W. Lu for their interest 
in this calculation. 
This work is supported in part by funds provided by the
U.S.  Department of Energy (D.O.E.) under cooperative agreement
DOE-FG02-93ER-40762.

\end{document}